# Electrically Driven Thermal Infrared Metasurface with Narrowband Emission

Xiu Liu, Lin Jing, Xiao Luo, Bowen Yu, Shen Du, Zexiao Wang, Hyeonggyun Kim, Yibai Zhong, Sheng Shen[*]

Department of Mechanical Engineering, Carnegie Mellon University, Pittsburgh, Pennsylvania 15213, USA

sshen1@cmu.edu

**ABSTRACT**

Metasurfaces consisting of an array of planar sub-wavelength structures have shown great potentials in controlling thermal infrared radiation, including intensity, coherence, and polarization. These capabilities together with the two-dimensional nature make thermal metasurfaces an ultracompact multifunctional platform for infrared light manipulation. Integrating the functionalities, such as amplitude, phase (spectrum and directionality), and polarization, on a single metasurface offers fascinating device responses. However, it remains a significant challenge to concurrently optimize the optical, electrical, and thermal responses of a thermal metasurface in a small footprint. In this work, we develop a center-contacted electrode line design for a thermal infrared metasurface based on a gold nanorod array, which allows local Joule heating to electrically excite the emission without undermining the localized surface plasmonic resonance. The narrowband emission of thermal metasurfaces and their robustness against temperature nonuniformity demonstrated in this work have important implications for the applications in infrared imaging, sensing, and energy harvesting.

Thermal infrared radiation emitted from materials with finite temperatures is ubiquitous but generally inefficient in applications, such as energy conversion and infrared sensing, due to its isotropic and incoherent characteristics. Recent advances in metasurfaces composed of an array of planar sub-wavelength scatters have provided new potentials for controlling thermal emission beyond these fundamental constraints [1–3]. The coupling among the thermal scatters or infrared antennas in metasurfaces can significantly modify emission spectrum[4–7], directionality[8–10], and polarization[11,12]. The low-cost metasurface-based thermal emitters can thus be excellent alternatives to semiconductor-based infrared light sources, which commonly require complicated cooling and band structure engineering[13–16].

Moreover, thermal infrared metasurfaces have emerged as an ultracompact multifunctional platform for light manipulation[17–19]. The two-dimensional nature of metasurfaces together with their strong capabilities in controlling thermal radiation makes them ideal for realizing flat, miniaturized and energy-saving on-chip infrared devices. Integrating the functionalities, such as amplitude, phase (spectrum and directionality), and polarization, on a single metasurface offers fascinating device responses[20,21], thus opening new opportunities for applications in lighting[22–24], sensing[25], imaging[26], and energy harvesting[27,28].

However, implementing these multifunctional optics in a small footprint generally involves many tradeoffs in optical, electrical, or thermal performances. For example, the finite-size and aspect-ratio effects hinder coupling between sub-wavelength structures on metasurfaces[29]. The high response speed resulting from heat conduction to the substrate can undermine thermal emission efficiency[30]. Moreover, the local Joule heating of metasurfaces typically needs infrared-transparent electrodes[31,32], which have limited choices and make the fabrication more complex. In this work, we develop an electrically driven thermal infrared metasurface based on a gold nanorod array that can achieve a narrowband nearly perfect emission through localized surface plasmonic resonance. The overall device, as shown in Fig. 1(a), is a metal-insulator-metal metasurface with a center-contacted electrode line design, which allows a local Joule heating to electrically excite the emission of the nanorods without deteriorating its optical narrowband resonance. This mechanism is justified by the optical simulations with and without the electrode lines.

Starting with a $SiO_2$ (1 μm thick)/Si substrate, we pattern a 100 nm Au layer with a rectangular shape that covers the whole metasurface active area as an underneath mirror for optical reflection but avoids overlapping with the top trapezoidal electrodes for the convenience of probing, as shown in the cross-section of Fig. 1(a). A 100 nm $Al_2O_3$ spacer between the Au mirror and the metasurface layer is then deposited by atomic layer deposition (ALD). Afterwards, by electron beam lithography, the metasurface pattern, consisting of a nanorod array where each column is connected by an electrode line, is patterned right above the mirror area, and is then transferred by liftoff after the deposition of 50 nm thick Au (with a 5 nm thick Cr adhesive layer) by electron beam evaporation. By tuning the gold nanorod dimensions, we can obtain the emission at different resonant wavelengths. The scanning electron microscope (SEM) images are shown in Fig. 1(b-c), indicating two fabricated nanorod arrays with emissivity resonance wavelengths at 5.21 μm and 3.54 μm, respectively. The dimensions of the metasurface with the resonance at 5.21 μm are periodicities $P_x = 1.8$ μm, $P_y = 0.5$ μm, length $L_x = 1.3$ μm, width $W_y = 0.15$ μm, and width of electrode line $W_{ce} = 0.1$ μm. The dimensions of the other one with the resonance at 3.54 μm are kept the same except $P_x = 1.3$ μm, $L_x = 0.8$ μm. To enable Joule heating, two trapezoidal electrodes, by sputtering 20 nm Cr, 180 nm Au, and 20 nm Pt in sequence, are aligned to connect the electrode line by photolithography. Finally, a very thin layer of $Al_2O_3$ (10 nm), deposited by ALD, covers the entire device to increase the heat tolerance of the fabricated gold nanorods[33].

The center-contacted electrode line design is verified by Ansys Lumerical finite-difference time-domain simulations. The entire device is illuminated by a plane wave source polarized along the long axis (x-axis) of the nanorods. A reflective power monitor is placed above the source to capture the reflective energy and thus the resonance feature of the gold nanorod array. As shown in Fig. 2(a), there is little difference in the reflectance between the nanorod arrays with and without the electrode lines for both the 5.21 μm and 3.54 μm resonances. Taking the nanorod array with the 5.21 μm resonance as an example, the minimal influence of the electrode lines can be illustrated in the electric field profiles shown in Fig. 2(b). The electrode line only occupies the area of the weakest field intensity adjacent to the center of the nanorod so that the nanorod shows an almost identical field distribution to the one without the electrode line. We can expand or shrink the electrode line freely to adjust impedance for a better power injection although the resonance can be undermined if the contacting size is larger than the weak-field area. This approach also keeps the metasurface with the simple physics of nanorod plasmonic resonators, whereby we can tune the narrowband emission straightforwardly through adjusting the nanorod dimension and the array periodicity, compared to the void metasurface designs[34–36].

We then check the temperature uniformity of the metasurface when it is electrically heated by the center-contacted electrode lines. Temperatures of the metasurface are measured by an emissivity-compensated thermal mapping. A 125 °C global heating is firstly applied to measure the emissivity of the unpowered metasurface. We then electrically drive the metasurface via a square pulse train at 2600 Hz with a duty cycle of 50%, and obtain average temperatures $T_{avg}$, via Stefan-Boltzmann's law, through a well-calibrated objective lens and an infrared camera. Here, voltage V is defined as on-state peak voltage while off-state is set as 0 V. Such Joule heating conditions of a square pulse train are also used for our later direct emission measurement where we need a modulated input for a lock-in Fourier transform infrared spectrometer (FTIR) setup[37].

In Fig. 3(a), the inset shows an infrared image of the array with a uniform hot area, along with which the average temperature $T_{avg}$ (average in the black-dashed area) and the spatial standard deviation of temperature (error bars) are also plotted. With the increased voltage V, the $T_{avg}$ increases parabolically corresponding to the Joule heat generation of $V^2/R$ where R is the electrical resistance. The temperature standard deviation, indicating the temperature non-uniformity, also increases with temperatures. A higher temperature tends to make the parabolic shape of temperature distribution from the Joule heating more prominent. For the metasurface with a size of $50 \times 50$ μm$^2$, the temperature profile demonstrates good uniformity in the center range of $40 \times 40$ μm$^2$. As shown in Fig. 3(b), the temperature profiles of the metasurfaces with voltage $V = 4$ V show that the temperature maintains uniform in the central 40 μm along both the x- and y-axes (defined in the inset of Fig. 3(a)). This uniformity is good enough to approximate Kirchhoff's law, which is also verified by our following direct emission measurement.

The narrowband direct emission of the metasurface is measured by our lock-in FTIR system, which combines a lock-in amplifier with the FTIR for noise reduction. Under a global heating temperature $T_0$, we apply the same square pulse train for the Joule heating as the one for the thermal mapping. The average temperature $T_{avg}$ is then generated from both the Joule heating by the voltage V and the global heating temperature $T_0$. The measured spectral signal is

$$S_{out}(\lambda, T_{avg}) = \varepsilon(\lambda)m(\lambda)I_{BB,m}(\lambda, T_{avg})$$

(1)

where $\varepsilon(\lambda)$ is the emissivity of the metasurface, $m(\lambda)$ is the wavelength-dependent response function from the Mercury-Cadmium-Telluride (MCT) detector and other optical components, and $I_{BB,m}(\lambda, T_{avg})$ is the modulated portion of blackbody emission power from the Joule heating[37,38].

The emissivity $\varepsilon(\lambda)$ of the metasurface can be obtained from Kirchhoff's law that equates the emissivity with the absorptivity (of the surface), which is the absorptance of incident radiation that can be measured from the indirect FTIR reflectance measurements via energy conservation. As shown in Fig. 4(a), the emissivity resonances of 5.21 μm and 3.54 μm are consistent with the indirect reflectance simulations in Figure 2, except that the peak values are halves since we use an unpolarized source to measure the reflectance. There is a higher-order mode for the emissivity resonance of 5.21 μm at about 3 μm, and these mismatches may attribute to the error in fabrication.

The measured direct emission spectra $S_{out}(\lambda, T_{avg})$ are shown in Fig. 4(b). It is worth to note that there is an obvious red shift of the direct emission resonance of 3.65 μm (dash-dot blue line) as compared to the indirect emissivity resonance of 3.54 μm (dashed black line). For a better comparison, we normalize all spectra and focus on their peak shifts in Fig. 4(c). With the $T_{avg}$ of 159.2 °C (generated by the global heating at $T_0 = 125$ °C and the Joule heating at $V = 5.0$ V), $S_{out}$ of the metasurface has an emission resonance around 3.65 μm with an about 0.11 μm red shift, compared to the emissivity resonance of 3.54 μm. The red shift reduces to about 0.06 μm, as indicated by $S_{out}$ with an emission resonance near 3.60 μm, when we increase $T_0$ up to 150 °C but keep the same V of 5.0 V resulting in $T_{avg}$ of 186.6 °C. Thus, we can conclude that the red shift results from the blackbody emission $I_{BB,m}(\lambda, T_{avg})$, which, under a relatively low temperature, has a peak position at a far longer wavelength than the emissivity resonance. The emission resonance of $S_{out}$, as the product of $\varepsilon$ and $I_{BB,m}$, is then dragged to a longer wavelength. Similarly, the emissivity resonance of 5.21 μm should also undergo a peak shift for its direct emission, but because of its better overlapping with the blackbody emission, the red shift turns out to be negligible. The wavelength-dependent response function $m(\lambda)$ may also contribute to this red shift, and we can specify $m(\lambda)$ with a known reference sample[39]. Another result from the non-overlapping between the emissivity $\varepsilon$ and the blackbody emission $I_{BB,m}$ is the peak intensity of the direct emission $S_{out}$. As shown in Fig. 4(b), even at a low $T_{avg}$ of 139.6 °C, the metasurface with the 5.21 μm emission resonance has a higher emission peak intensity than the one with the emission resonance of 3.65 μm, whose $T_{avg}$ is 159.2 °C, due to its better overlapping with the peak of blackbody emission spectrum of 139.6 °C.

Overall, the direct emission of the metasurface shows a good spectral coherence and matches the emissivity resonances in Fig. 4(a). Such the good consistency indicates the robustness of our thermal infrared metasurface to the temperature non-uniformity since only $40 \times 40$ μm$^2$ of $50 \times 50$ μm$^2$ exhibits uniform temperature under our Joule heating conditions, as discussed in Figure 3. On the other hand, the non-uniformity in temperature can be included via a local Kirchhoff's law[40] for a more accurate direct emission spectrum. The true emission power can also be computed if the system response $m(\lambda)$ is calibrated [37].

In summary, the center-contacted electrode line design described in this work allows an electrically driven thermal infrared metasurface to demonstrate the simple physics of nanorod plasmonic resonators with a polarized nearly perfect narrowband emission. The resonances can be easily tuned by changing the dimensions and periodicity of the nanorods. The temperature uniformity shown in this work can be further improved by optimizing the substrate design[29]. The narrowband direct emission shows clear spectral coherence, consistent with the indirect reflectance measurement. The idea of contacting the weak-field area for the Joule heating can be extended to other metasurface geometries, such as crossbars with some non-polarized emission. Also, the electrode lines only heat up the nanorod center areas and can potentially benefit a high-speed modulation[41]. Thus, this special design of electrically driven metasurface is expected to contribute to the pursuit of an ultracompact multifunctional infrared light manipulation.

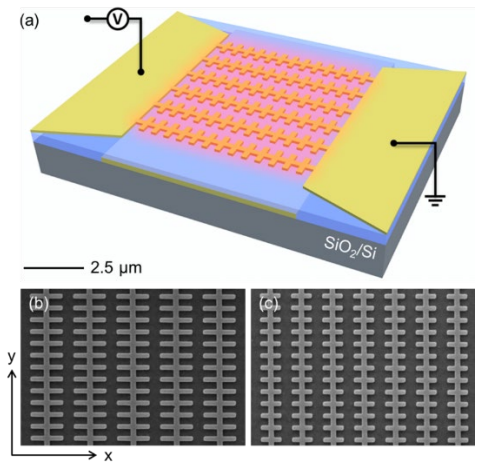

**FIG. 1.** (a) Schematic of an electrically driven thermal infrared metasurface. (b) SEM image of the gold nanorod array with resonance at 5.21 μm and its centered-contact electrode lines. (c) SEM image of the metasurface with the same design but resonating at 3.54 μm. The optical resonances are mainly dependent on the nanorod dimension and the array periodicity. The center-contacted electrode lines work as electrodes for the Joule heating without undermining the narrowband emission.

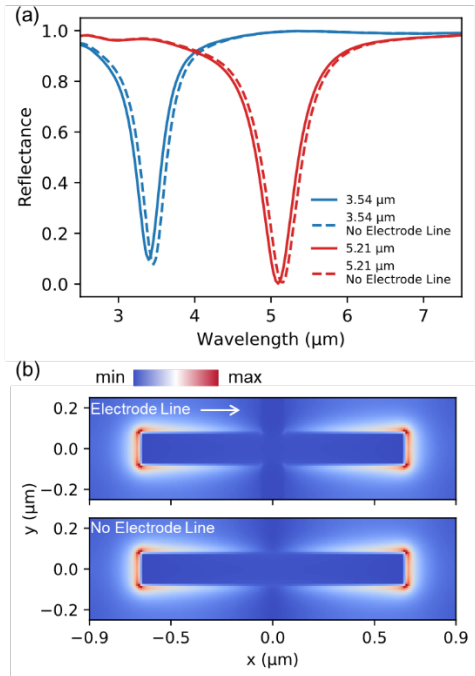

**FIG. 2.** (a) Reflectance of metasurfaces with and without electrode lines for both 5.21 μm and 3.54 μm resonances. (b) Electric field profiles of nanorod in one period at 5.21 μm resonance with and without electrode lines. Since the electrode lines contact the center area of the nanorod where the field intensity is the weakest, the overall field distribution keeps almost identical to the one without the electrode lines. Thus, there is little difference between the reflectance with electrode lines and the one without electrode lines.

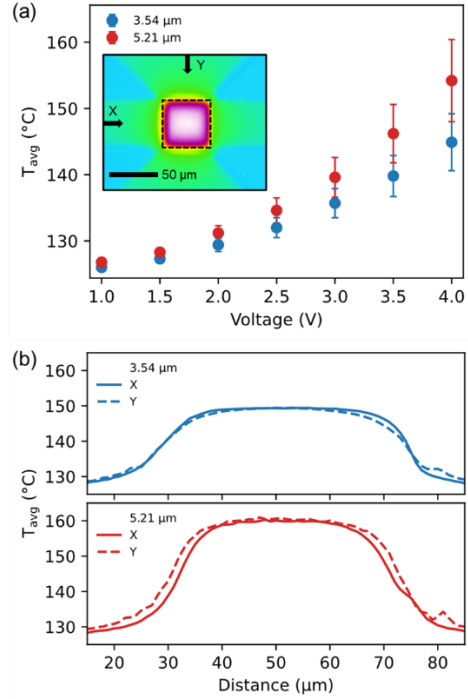

**FIG. 3.** (a) Average temperatures $T_{avg}$ of metasurface at different voltages V. The inset shows an emission image of the array. (b) Average temperature profiles of metasurfaces along x- and y-axes for resonances of 3.54 μm and 5.21 μm, respectively. With increasing V, the $T_{avg}$ increases parabolically following the rule of the Joule heating, and the temperature uniformity becomes worse but is still enough to approximate Kirchhoff's law that equals emissivity to absorptivity of the metasurfaces.

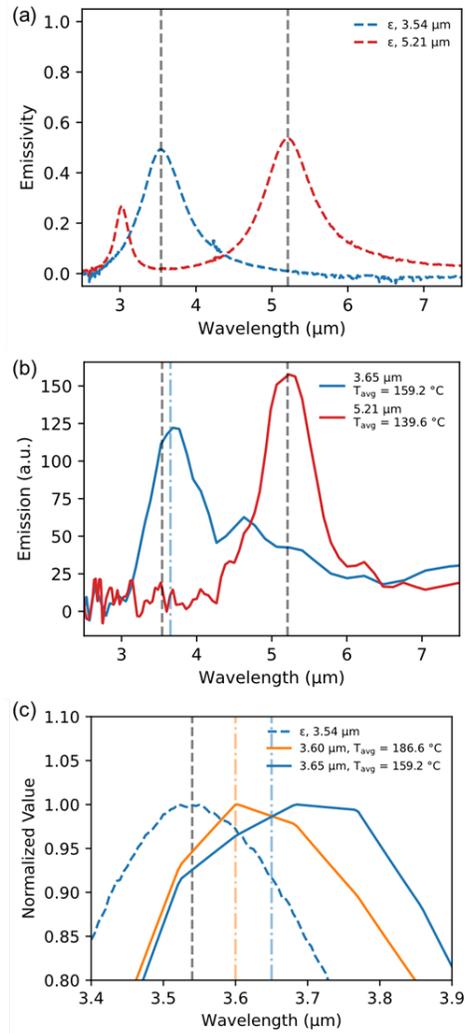

**FIG. 4.** (a) Indirect emissivity spectra of metasurfaces for both 5.21 μm and 3.54 μm resonances (b) Direct emission spectra of metasurfaces for both 5.21 μm and 3.65 μm resonances. (c) Normalized spectra for a comparison of emissivity resonance. The direct emission spectra of the metasurfaces show a nice spectral coherence with their resonances consistent with the indirect emissivity spectra, except a small red shift due to the non-overlapping between the emissivity and the corresponding blackbody emission.